\documentclass[preprint]{revtex4}

\usepackage{graphicx}
\usepackage{amsmath}
\usepackage{amssymb}

\newcommand{\cA}{\varphi_1}
\newcommand{\cB}{\varphi_2}
\newcommand{\cC}{\varphi_3}
\newcommand{\cD}{\varphi_4}
\newcommand{\cmc}{\varphi_{\rm cmc}}
\newcommand{\const}{{\mbox{const}}}
\newcommand{\cnuc}{c_{\rm nuc}}
\newcommand{\eg}{e.g., }
\newcommand{\ie}{i.e., }
\newcommand{\kT}{k_{\rm B}T}
\newcommand{\ldif}{\ell_{\rm d}}
\newcommand{\mhat}{\hat{m}}
\newcommand{\mkin}{\bar{m}}
\newcommand{\mmax}{m_{\rm nuc}}
\newcommand{\mmin}{m^{*}}
\newcommand{\pd}{\partial}
\newcommand{\Phibb}{\Phi_1^{\rm b}}
\newcommand{\Phib}{\Phi_{\rm b}}
\newcommand{\Phikin}{\bar{\Phi}_1}
\newcommand{\Phimin}{\Phi_1^{*}}
\newcommand{\Phinuc}{\Phi_{\rm nuc}}

\newcommand{\psihat}{\hat{\psi}}
\newcommand{\rme}{{\rm e}}
\newcommand{\taudif}{\tau_{\rm d}}
\newcommand{\taukin}{\tau_{\rm k}}
\newcommand{\taunuc}{\tau_{\rm nuc}}
\newcommand{\taurel}{\tau_{\rm r}}

\begin{document}
\author{Radina Hadgiivanova}
\affiliation{Present address: Department of Chemistry \& Biochemistry,
University of California, Santa Barbara, CA 93106-9510}
\author{Haim Diamant}
\email{hdiamant@tau.ac.il}
\affiliation{Raymond and Beverly Sackler School of Chemistry,
Tel Aviv University, Tel Aviv 69978, Israel}
\author{David Andelman}
\affiliation{Raymond and Beverly Sackler School of Physics and Astronomy,
Tel Aviv University, Tel Aviv 69978, Israel}

\title{Kinetics of Surfactant Micellization:\\ A Free Energy Approach}

\date{Nov 30, 2010}

\baselineskip=12pt 
\begin{abstract}

We present a new theoretical approach to the kinetics of micelle
formation in surfactant solutions, in which the various stages of
aggregation are treated as constrained paths on a single free-energy
landscape. Three stages of well-separated time scales are
distinguished. The first and longest stage involves homogeneous
nucleation of micelles, for which we derive the size of the critical
nuclei, their concentration, and the nucleation rate. Subsequently, a
much faster growth stage takes place, which is found to be
diffusion-limited for surfactant concentrations slightly above the
critical micellar concentration ({\it cmc}), and either
diffusion-limited or kinetically limited for higher concentrations.
The time evolution of the growth is derived for both cases.  At the
end of the growth stage the micelle size may be either larger or
smaller than its equilibrium value, depending on concentration.  A
final stage of equilibration follows, during which the micelles relax
to their equilibrium size through fission or fusion. Both cases of
fixed surfactant concentration (closed system) and contact with a
reservoir of surfactant monomers (open system) are addressed and found
to exhibit very different kinetics. In particular, we find that
micelle formation in an open system should be kinetically suppressed
over macroscopic times and involve two stages of micelle nucleation
rather than one.

\end{abstract}

\maketitle

\section{Introduction}
\label{sec_intro}

Self-assembly of amphiphilic molecules into mesoscopic
(micelles) in solution is a common and thoroughly investigated
phenomenon \cite{Israelachvili}. Dynamic aspects of this process have
been extensively studied as well \cite{Zana_book}. The techniques
applied in such experiments and the interpretation of their results
have used the framework of reaction kinetics, where each aggregate
size is treated as a distinct chemical species, and changes in size
and population --- as chemical reactions (ref \citenum{Zana_book},
chapter 3). Two well-separated time scales are identified in
experiments \cite{Zana76}. The shorter of the two, denoted $\tau_1$
(typically $\sim 10^{-5}$--$10^{-4}$ s), corresponds to the exchange
of a single molecule between a micelle and the monomeric solution;
during this time scale the number of micelles remains essentially
fixed. The second time scale, $\tau_2$ (which widely varies and may
be, \eg about $\sim 10^{-2}$ s) is associated with overcoming the
barrier to the formation or disintegration of an entire micelle. The
total activation time for such a process is $m\tau_2$, where $m$ is
the number of molecules in a micelle.  During this longer time scale
the number of micelles changes.

The first and still prevalent theory of micellar kinetics by Aniansson
and Wall \cite{AW} is based on these observations.  It casts the
micellization process in the form of reaction kinetics with two well
separated time scales, whereby micelles form and disintegrate through
a series of single monomer-exchange reactions. While various
extensions to the Aniansson-Wall theory have been presented over the
years
\cite{Almgren,Lessner,Kahlweit,Hall,Wall,Aniansson85,DeMaeyer,Rusanov,Semenov},
only a few alternative approaches have been suggested.  In ref
\citenum{Ball} the interesting possibility that micellization may
behave as a bistable autocatalytic reaction was explored. An idealized
(one-dimensional) nucleation model for linear aggregates was suggested
in ref \citenum{Neu}. An important alternative approach to study
micellization kinetics has been the use of computer simulations
\cite{simulation1,simulation2,simulation3,simulation4,simulation5,simulation6,simulation7,simulation8}.
In the case of micellization of amphiphilic block copolymers more
progress has been achieved (ref \citenum{Zana_book}, chapter 4; refs
\citenum{Halperin}--\citenum{Bates}). The kinetics of such polymeric
micelles, however, usually depends on qualitatively different effects
--- in particular, the high entropy barrier for polymer penetration
into a micelle.

In the current work we present a new approach to the kinetics of
surfactant micellization, which is based on a free-energy formalism. A
similar strategy was previously applied to the kinetics of surfactant
adsorption at interfaces \cite{jpc96,csa01}. This approach has two
main advantages. The first is that it provides a more unified
description of the kinetics --- rather than considering different
stages as separate processes (``reactions''), they can all be cast as
constrained pathways on a single free-energy landscape.  Considering
different processes on the same footing allows, for example, easier
identification of rate-limiting stages such as diffusion-limited or
kinetically limited ones \cite{csa01}. The second advantage of such a
formalism is that it can be relatively easily extended to more complex
situations, such as ionic solutions or surfactant mixtures
\cite{Lang99}. On the other hand, the shortcoming of the model is that
it is phenomenological, following coarse-grained thermodynamic
variables rather than those characterizing single molecules and
aggregates. It is probably not appropriate for large polymeric
micelles, where intra-chain degrees of freedom play an important role
and a more detailed description of molecules and aggregates is
required \cite{Halperin,Besseling,Zana06}.  We shall focus here,
therefore, on the micellization of short-chain surfactants.

Another consequence of the coarse-grained modeling is that the
derivation for the kinetics of micellization bears similarities to the
kinetics of first-order phase transitions --- an analogy that was
previously invoked \cite{Neu,Besseling}. However, unlike macroscopic
phase separation, micellization is restricted to finite-size
aggregates, resulting, for example, in growth laws that are not
scale-free.

In the next section we present the free-energy formalism and its
implications for the process of micelle formation. As in previous
theories we subsequently separate the kinetics into stages of
disparate time scales, during each of which a different set of
constraints is imposed. We discuss separately the kinetics of closed
and open systems.  A closed system contains a fixed number of
surfactant molecules. In an open system the surfactant solution is in
contact with a large reservoir, which is at thermodynamic equilibrium.
Whereas in equilibrium this distinction is usually immaterial, the
kinetics of the two cases are found to be strikingly different. While
reading through the various stages of micellization it may be helpful
to refer to the two schematic diagrams provided at the end of the
article (\ref{fig_chart1} and \ref{fig_chart2} for closed and open
systems, respectively). The first stage that we address is the
nucleation of micelles. Subsequently, we describe the growth of the
micellar nuclei as they absorb additional monomers from the
surrounding solution. Both options of kinetically limited and
diffusion-limited growth are studied. In addition, the possible role
of long-distance diffusive transport is examined.  We then consider
the final relaxation toward equilibrium. Finally, we summarize the
conclusions and discuss the experimental implications of our analysis,
as well as its limits of validity.

\section{Model}
\label{sec_model}

The model is based on a simple free-energy functional, which has been
recently introduced to study metastability issues of micellization
\cite{jpcb07}. Apart from the temperature $T$, the free energy depends
on three thermodynamic degrees of freedom, which we take to be the
total volume fraction of surfactant in the solution, $\Phi$, the
volume fraction of surfactant monomers, $\Phi_1$, and the number of
molecules in a micelle (aggregation number), $m$.  Despite the
simplified two-state (monomer--aggregate) description, polydispersity
can be accounted for as fluctuations of the variable $m$
\cite{jcp09}. (This, however, restricts the validity of the model to
compact micelles whose size distribution is narrow
\cite{Israelachvili}.) All energies hereafter are given in
units of the thermal energy, $k_{\rm B}T$.

The free energy has contributions from the entropy of mixing and from
the interactions among surfactant molecules. The former is obtained
from a coarse-grained lattice scheme (Flory-Huggins model), in which a
water molecule occupies a single lattice cell of volume $a^3$, and a
surfactant molecule is larger and occupies $n$ such cells. The
interactions in the solution are represented by a single
phenomenological function, $u(m)$, which is assumed to capture all the
molecular contributions to the free energy of transferring a
surfactant molecule from the solution into an aggregate of size
$m$. The resulting Helmholtz free energy density (per lattice site) is
\cite{jpcb07}
\begin{equation}
  F(\Phi,\Phi_1,m) = \frac{\Phi_1}{n}\ln\Phi_1
  + \frac{\Phi_m}{nm} [\ln\Phi_m - mu(m)]
  + (1-\Phi)\ln(1-\Phi),
\label{F}
\end{equation}
where $\Phi_m=\Phi-\Phi_1$ is the volume fraction of micelles, and
$1-\Phi$ is the volume fraction of water.  At equilibrium the solution
is spatially uniform and characterized by those single mean values of
the variables, which minimize the free energy under the appropriate
constraints. For a closed system $F$ is minimized with respect to
$\Phi_1$ and $m$ for a given $\Phi$. For an open system one should
minimize $F-\mu\Phi$ with respect to $\Phi$, $\Phi_1$, and $m$ for a
given surfactant chemical potential $\mu$.  Out of equilibrium the
values of variables, such as $\Phi$, $\Phi_1$, and $m$, may be
position-dependent, and the total free energy is given by spatial
integration of the local free-energy density.  (We neglect here
surface-tension (gradient) terms associated with boundaries between
such spatial domains.)

The specific choice of the interaction function $u(m)$ is not crucial
so long as it has a maximum at a finite $m$ to ensure the stability of
finite-size micelles. To provide numerical examples, and following
previous works \cite{jpcb07,Chandler}, we use a simple three-parameter
function,
\begin{equation}
  u(m) = u_0 - \sigma m^{-1/3} - \kappa m^{2/3}.
\label{u}
\end{equation}
The first term in eq \ref{u} represents a
micelle-size-independent free-energy gain in increasing $m$, the
second --- a surface energy penalty, and the third is responsible for
stabilizing a finite-size aggregate.  (For a more detailed discussion
of these terms and the restricted ranges of relevant values for $u_0$,
$\sigma$, and $\kappa$, see ref \citenum{jpcb07}.)

\begin{figure}
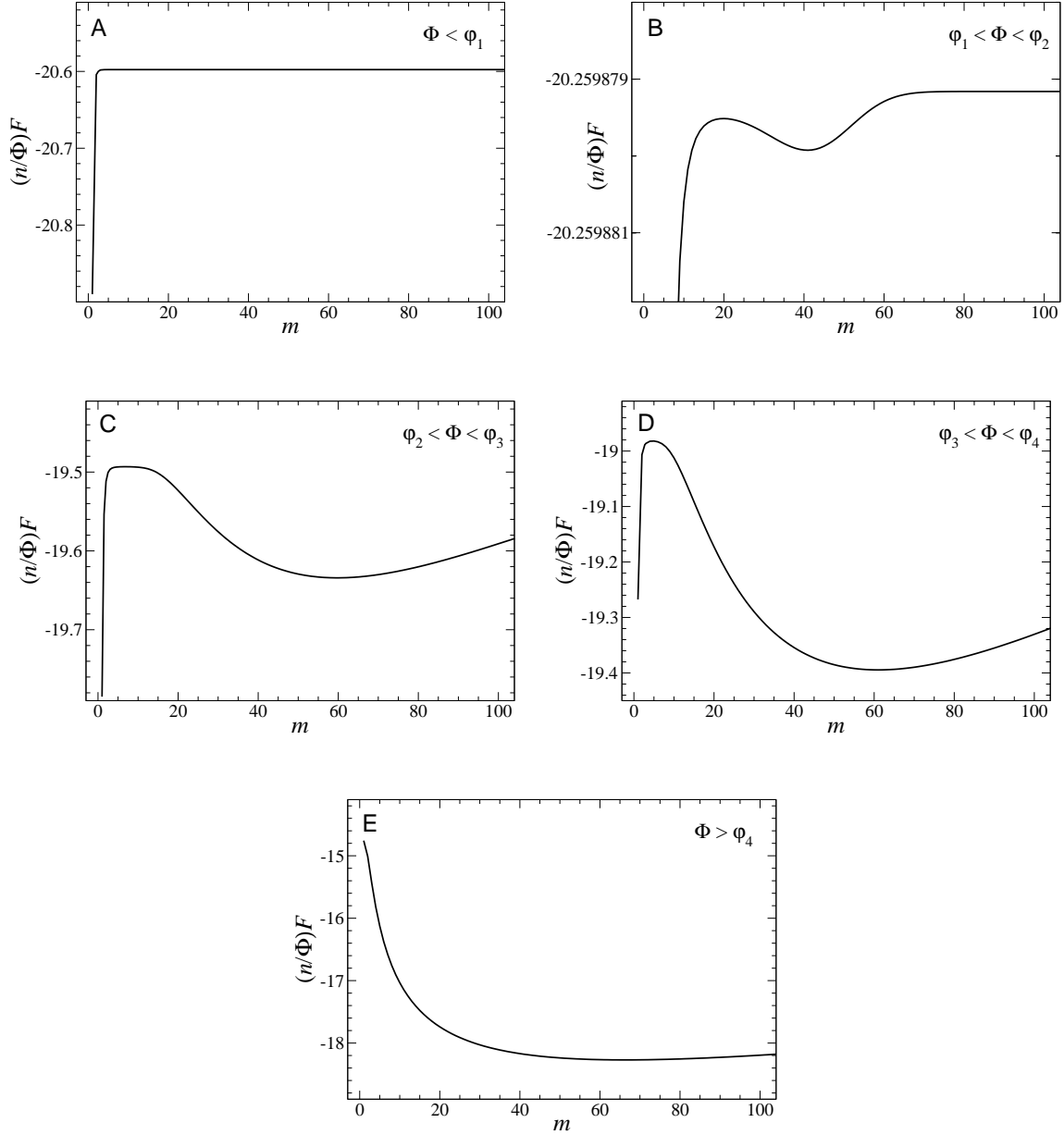

\includegraphics[height=4.8cm]{fig1a.eps}
\hspace{0.3cm}
\includegraphics[height=4.8cm]{fig1b.eps}\\
\vspace{0.9cm}
\includegraphics[height=4.8cm]{fig1c.eps}
\hspace{0.3cm}
\includegraphics[height=4.8cm]{fig1d.eps}\\
\vspace{0.9cm}
\includegraphics[height=4.8cm]{fig1e.eps}
\caption{Cuts of the free-energy landscape (per surfactant molecule,
  in units of $\kT$) as a function of aggregation number along the
  $\Phimin(m)$ line for the surfactant parameters of
  \ref{tab_surfactant} and increasing surfactant volume fraction,
  $\Phi$: (A) $\Phi=5\times 10^{-4}<\cA$; (B) $\cA<\Phi=7\times
  10^{-4}<\cB$; (C) $\cB<\Phi=1.5\times 10^{-3}<\cC=\cmc$; (D)
  $\cC<\Phi=2.5\times 10^{-3}<\cD$; and (E) $\Phi=0.11>\cD$.}
\label{fig_landscape}
\end{figure}

Despite its simplicity eq \ref{F} defines a rather rich
free-energy landscape over a three-dimensional space of macro-states,
parametrized by the axes $(\Phi,\Phi_1,m)$. We are going to treat the
kinetics of micellization as time-dependent paths along this
landscape, and it is beneficial, therefore, to first recall its key
features \cite{jpcb07}, demonstrated in \ref{fig_landscape}. For any
given $\Phi$ and along the $\Phi_1$ axis, $F$ always has a single
minimum at $\Phi_1=\Phimin(m,\Phi)$ for all values of $m$. This value
of monomer volume fraction as a function of aggregation number and
total volume fraction is found by solving the equation
\begin{equation}
  \Phi_1=\Phimin(m,\Phi):\ \ \ (\Phi_1)^m \rme^{mu(m)+m-1} = \Phi-\Phi_1.
\label{phimin}
\end{equation}
Along the $m$ axis, however, $F$ becomes nonconvex when $\phi$ exceeds
a certain volume fraction, $\cA$, with two minima at $m=1$ and
$m=\mmin(\Phi_1,\Phi)$, and a maximum in between at
$m=\mmax(\Phi_1,\Phi)$. (See \ref{fig_landscape} A and B.) The extrema
satisfy the equation
\begin{equation}
  m=\mmin,\mmax:\ \ \ m^2 = -\ln(\Phi-\Phi_1)/u\,'(m),
\label{mextreme}
\end{equation}
where $u\,'=du/dm$.  Combining eqs \ref{phimin} and
\ref{mextreme}, we can find $m$ and $\Phi_1$ at the extrema for a
given $\Phi$ according to
\begin{eqnarray}
  m=\mmin,\mmax:\ \ \ m^2 &=& -\ln[\Phi-e^{-u(m)-mu\,'(m)-1+1/m}]/u\,'(m),
 \label{mextreme2} \\
  \Phimin &=& e^{-u(m)-mu\,'(m)-1+1/m}.
 \label{phi1extreme}
\end{eqnarray}

Above a larger volume fraction, $\cB>\cA$ (\ref{fig_landscape}C), the
micellar state with $\Phi>\cB$, $m=\mmin$ and
$\Phi_1=\Phimin(\mmin,\Phi)$, though still metastable, may become
appreciably occupied, giving rise to premicellar aggregates
\cite{jpcb07}. Above yet another volume fraction, $\cC>\cB$
(\ref{fig_landscape}D), the micellar state for $\Phi>\cC$ becomes the
global minimum of $F$.  It is this point, analogous to the binodal
line in phase separation, which corresponds to the commonly defined
{\it cmc} \cite{jpcb07}, \ie ${\it cmc}=c_3=\cC/(na^3)$. We shall
focus in the current work on the ordinary micellization region,
$\Phi>\cC=\cmc$, where micelles are stable at equilibrium. It should
be kept in mind, however, that in this region the monomeric and
micellar states are separated by a free-energy barrier in the form of
a saddle point of $F$, $F_{\rm
nuc}(\Phi)=F[\Phi,\Phimin(\mmax,\Phi),\mmax]$. The barrier may be
high, leading to the measurement of an apparent {\it cmc}, which is
higher than the equilibrium one, $\cC=\cmc$ \cite{Semenov}. Finally,
above a certain higher volume fraction, $\cD>\cC$, the barrier
disappears and the micellar state for $\Phi>\cD$ remains the sole
minimum of $F$, as seen in \ref{fig_landscape}E. (This work is
restricted to the isotropic micellar phase of surfactant solutions; at
higher concentrations other phases and meso-phases appear
\cite{Israelachvili}.) The point $\Phi=\cD$ is the analogue of
the spinodal line in macroscopic phase separation.  As already
mentioned in the Introduction, despite the analogy with phase
separation it should be borne in mind that micellization is
essentially different in that it involves finite-size aggregates and
smooth crossovers rather than macroscopic phases and sharp
transitions.

The initial and final states of the micellization kinetics are defined
as follows. At $t=0$ the system is in the monomeric state,
$(\Phi_1=\Phi,m=1)$, whereas its equilibrium state is the micellar
one.  In a closed system this is done by setting the surfactant volume
fraction above the {\it cmc}, $\Phi>\cC$ (using, for example, the
temperature-jump or stopped-flow techniques \cite{Zana_book}). In an
open system the initial condition corresponds to opening a diffusive
contact with a bulk reservoir, whose surfactant volume fraction
$\Phib$ is above the {\it cmc}, $\Phib>\cC$. The reservoir is assumed
to have already reached the equilibrium micellar state. At
$t\rightarrow\infty$ the system reaches the global minimum of the free
energy --- $[\Phi,\Phimin(\mmin,\Phi),\mmin(\Phi)]$ in the closed case
and $[\Phib,\Phimin(\mmin,\Phib),\mmin(\Phib)]$ in the open one. In
what follows we consider the kinetic pathway that the system takes
between these initial and final states. Assuming separation of time
scales, we shall divide the temporal path into separate stages. Note
that the various time scales are derived from the free energy
functional and a single molecular time, $\tau_0$, thus enabling
comparison of different stages and processes.

Throughout the following sections we demonstrate the results using a
single exemplary surfactant, whose parameters are listed in
\ref{tab_surfactant}. This allows comparison with refs
\citenum{jpcb07,jcp09}, where the behavior of the same exemplary
surfactant for $\Phi<\cC=\cmc$ was presented.

\begin{table}[tbh]
\caption{Parameters of the exemplary surfactant and the resulting
boundaries of the micellar region}
\begin{center}
\begin{tabular}{c c c c c c}
\hline
$n$ & $u_0$ & $\sigma$ & $\kappa$ & $\cC=\cmc$ & $\cD$ \\
\hline
13 & 10 & 11 & 0.08 & $2.03\times 10^{-3}$ & 0.106\\ 
\hline
\end{tabular}
\end{center}
\label{tab_surfactant}
\end{table}

\begin{figure}
\vspace{0.5cm}
\includegraphics[width=0.48\textwidth]{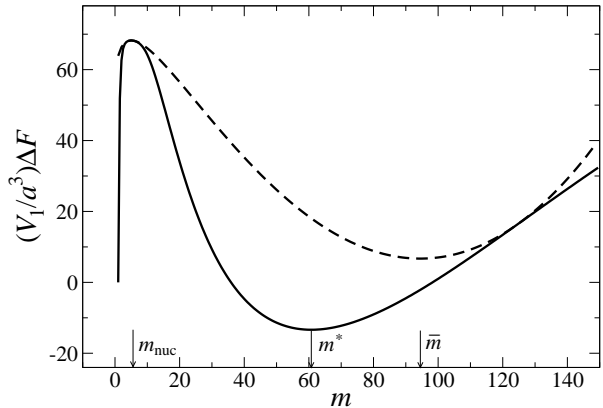}
\caption{Cuts of the free energy landscape (relative to the monomeric
state, per micelle, in units of $\kT$) as a function of aggregation
number for the surfactant parameters of \ref{tab_surfactant} and
$\Phi=1.1\cC$.  The two curves correspond to two different
constraints: relaxation of the monomer volume fraction for the given
aggregation number (solid), and fixed concentration of micelles
(dashed). Indicated by arrows are the sizes of the critical nucleus
($\mmax$), the intermediate aggregate at the end of the growth stage
($\mkin$), and the equilibrium micelle ($\mmin$). A closed system is
assumed. $V_1$ is the volume of solution per micelle at equilibrium.}
\label{fig_F}
\end{figure}

\ref{fig_F} shows two cuts through the free-energy landscape as
a function of aggregation number for the exemplary surfactant in a
closed system at total surfactant volume fraction slightly larger than
$\cC=\cmc$. Along the first cut (solid line) the monomer volume fraction is
assumed to be at quasi-equilibrium, $\Phi_1=\Phimin(m)$. Thus, the
minimum of this curve corresponds to the global minimum --- the
equilibrium aggregation number. Along the other cut (dashed curve),
which is relevant to the next two sections, we constrain the
concentration of micelles to remain at its nucleation value.

\section{Micellar Nucleation}
\label{sec_nucleation}

\subsection{Closed System}

Let us set the total volume fraction at $t=0$ to some value,
$\Phi>\cC=\cmc$, and assume that value (apart from a short initial period
of homogenization which is ignored) to remain fixed and uniform
throughout the micellization process. The first stage to consider is
the ascent of the free energy from the initial metastable state,
$(\Phi,\Phi_1=\Phi,m=1)$, to the saddle point
$[\Phi,\Phi_1=\Phimin(\mmax),m=\mmax]$ --- \ie the formation of the
critical nuclei. This activated process is assumed to be much slower
than diffusion. Hence, $\Phi_1$ can be taken during this stage as
spatially uniform and equal to the value that minimizes the free
energy for the given $\Phi$ and $m(t)$. Thus, as $m(t)$ increases from
$1$ to the critical-nucleus size $\mmax$, the system proceeds along
the path that satisfies the constraints $\Phi=\const$ and
$\Phi_1=\Phimin[m(t),\Phi]$.

A similarly constrained path was studied in detail in ref
\citenum{jcp09} to obtain the lifetime of metastable micelles in the
region $\cB<\Phi<\cC$ using Kramers' theory. Such a rigorous
calculation, unfortunately, cannot be repeated here, since the
metastable monomeric state is actually not a local minimum of $F$ but
just the edge, at $m=1$, of the range of allowed aggregation numbers.
(See \ref{fig_landscape}D.) Nevertheless, as demonstrated in ref
\citenum{jcp09}, the nucleation time (dissociation time in ref
\citenum{jcp09}) and its concentration dependence are primarily
determined by the height of the free-energy barrier.

The free-energy barrier corresponds to the nucleation of a single
micelle.  Our model, however, considers macrostates of a solution
containing many micelles and monomers. To switch between these two
descriptions we introduce a subsystem volume, $V_1$, which contains
(on average) a single nucleus. The volume fraction of critical nuclei,
their concentration, and the volume per nucleus are readily given for
closed systems by
\begin{eqnarray}
\Phinuc(\Phi) &=& \Phi - \Phimin[\mmax(\Phi),\Phi]
 \nonumber\\
  \cnuc(\Phi) &=& \Phinuc(\Phi) / [na^3\mmax(\Phi)] \nonumber\\
  V_1(\Phi) &=& \cnuc^{-1} = \frac{na^3\mmax(\Phi)}{\Phi - \Phimin[\mmax(\Phi),\Phi]},
\label{cnuc}
\end{eqnarray}
where $\mmax(\Phi)$ and $\Phimin(\Phi)$ are given by eqs
\ref{mextreme2} and \ref{phi1extreme}.  Since $\Phinuc$ is
very small, $V_1$ is much larger than the molecular volume, and our
coarse-grained approach is indeed applicable. Note the distinction
between the nuclei concentration $\cnuc$ and their volume fraction
$\Phinuc$. Since the micelle size $m$ is a variable, constraining
$\cnuc$ does not imply a fixed $\Phinuc$. This will be important in
the next sections, when we impose a constraint on the number of
nuclei. The nucleation barrier and nucleation time scale are given for
closed systems by
\begin{eqnarray}
  \Delta F_{\rm nuc}(\Phi) &=& \frac{V_1(\Phi)}{a^3}
  \left\{ F[\Phi,\Phimin(\mmax,\Phi),\mmax]
  - F_1(\Phi) \right\}\nonumber\\
  \taunuc(\Phi) &\simeq& \tau_0 \rme^{\Delta F_{\rm nuc}(\Phi)},
\label{taunuc}
\end{eqnarray}
where $\tau_0$ is a molecular time scale, and $F_1$ is the free energy
of the monomeric state. It should be mentioned that our formalism
artificially distinguishes between monomers and aggregates of size
$m=1$. As in the previous works \cite{jpcb07,jcp09}, this artifact has
an insignificant effect on the results. We calculate here the free
energy of the $m=1$ state as $F_1(\Phi)=F[\Phi,\Phimin(1,\Phi),1]$.

\begin{figure}
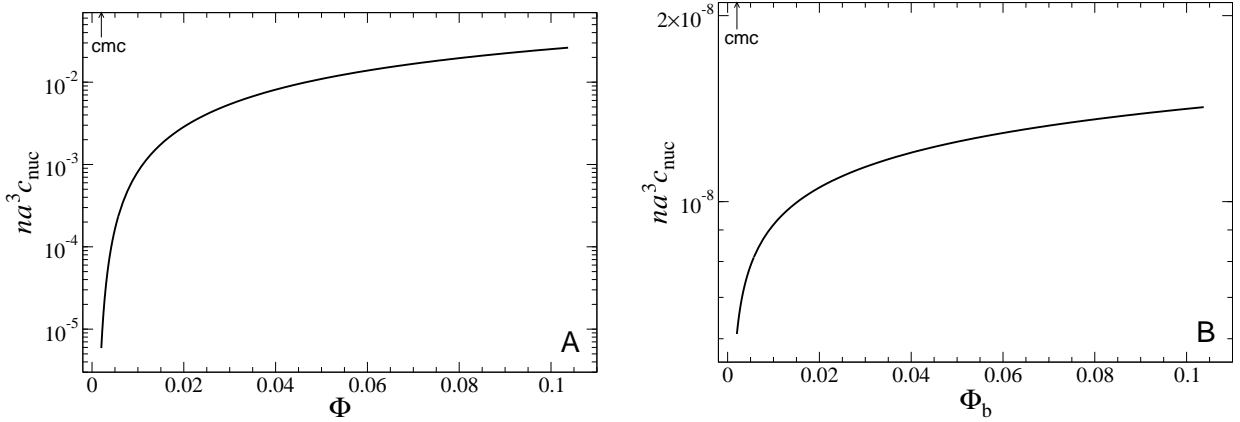

\vspace{0.8cm}
\includegraphics[width=0.48\textwidth]{fig3a.eps}
\hspace{.3cm}
\includegraphics[width=0.48\textwidth]{fig3b.eps}
\caption{Concentration of critical nuclei (normalized by the molecular
volume) as a function of surfactant volume fraction in the range
between $\cC=\cmc\simeq 2\times 10^{-3}$ and $\cD$ for closed (A) and open (B)
systems. Parameters are given in \ref{tab_surfactant}.}
\label{fig_cnuc}
\end{figure}

\begin{figure}
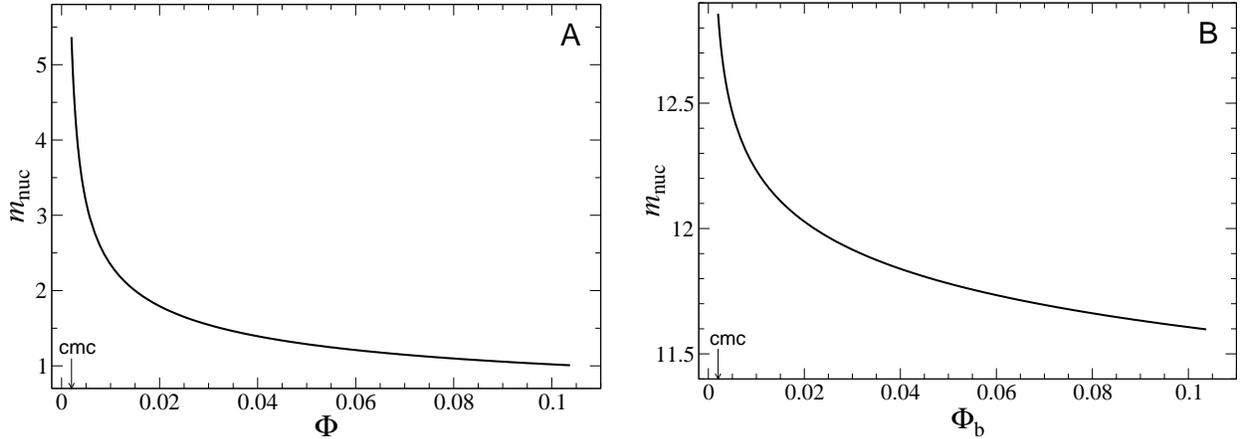

\includegraphics[width=0.48\textwidth]{fig4a.eps}
\hspace{.3cm}
\includegraphics[width=0.48\textwidth]{fig4b.eps}
\caption{Critical-nucleus size as a function of surfactant volume
  fraction in the range between $\cC=\cmc\simeq 2\times 10^{-3}$ and $\cD$
  for closed (A) and open (B) systems. Parameters are given in
  \ref{tab_surfactant}.}
\label{fig_mmax}
\end{figure}

\begin{figure}[tbh]
\vspace{0.8cm}
\includegraphics[width=0.48\textwidth]{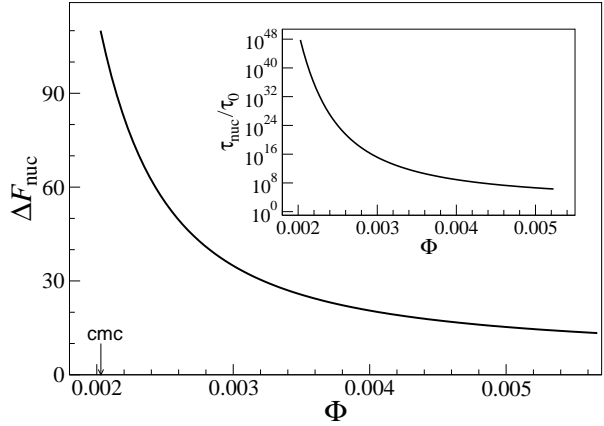}
\caption{Nucleation barrier $\Delta F_{\rm nuc}$ (in units of $\kT$)
  as a function of surfactant volume fraction for a closed system. The
  inset shows the corresponding nucleation time (in units of the
  molecular time $\tau_0$). Parameters are given in
  \ref{tab_surfactant}.}
\label{fig_Fnuc}
\end{figure}

Various features of the nucleation stage can be calculated from eqs
\ref{F}--\ref{taunuc}, as demonstrated in Figures
\ref{fig_cnuc}--\ref{fig_Fnuc}. The concentration of
critical nuclei (\ref{fig_cnuc}A) sharply increases with surfactant
volume fraction as $\Phi$ is increased above $\cC=\cmc$. The size of
the critical nucleus (\ref{fig_mmax}A) decreases with $\Phi$ until it
practically vanishes as $\Phi$ approaches $\cD$. The height of the
nucleation barrier (\ref{fig_Fnuc}) decreases as well with $\Phi$,
leading to a sharp decrease in the nucleation time scale
(\ref{fig_Fnuc} inset). To get an estimate of the actual nucleation
time scales we may take $\tau_0\sim 10^{-8}$ s, which is the time it
takes a molecule with a diffusion coefficient $\sim 10^{-6}$
cm$^2$~s$^{-1}$ to be displaced by $\sim 1$ nm. For the example
presented in \ref{fig_Fnuc}, $\taunuc$ is extremely large close to
$\cC$ but drops to $\sim 1$ s for $\Phi\simeq 2\cC$.

\subsection{Open System}
\label{sec_nuc_open}

When the system is placed in contact with a large reservoir of volume
fraction $\Phib$, $\Phib>\cC=\cmc$, monomers will first diffuse in,
until the monomeric concentrations are balanced. We shall assume that
micellar diffusion from the reservoir is either blocked or very slow.
(If it is not, micellization in the system will be dominated by simple
transport of micelles from the reservoir.) Thus, the starting point
for the nucleation stage in this case is different from that of a
closed system --- it is still a monomeric state, yet with a lower
volume fraction, $\Phi=\Phi_1$ where $\Phi_1=\Phibb<\Phib$.
Nucleation is again assumed much slower than monomer diffusion.
Hence, the monomer volume fraction remains fixed at $\Phi_1=\Phibb$.
At the same time it should minimize $F$ for the given $m(t)$, which in
turn determines the value of the third state variable, $\Phi$. As the
nuclei grow, the total volume fraction increases, and the system
proceeds along the path that satisfies the constraints $\Phi_1=\Phibb$
and $\Phimin[m(t),\Phi]=\Phibb$.

The nucleation path ends at the state of critical nuclei, which is
also different from the closed-system saddle point, because the total
volume fraction has not reached the bulk value, $\Phi<\Phib$. This
state is calculated using the following procedure for the open case.
First, we calculate the monomer volume fraction in the reservoir
according to the equilibrium condition,
\begin{equation}
  \Phibb(\Phib) = \Phimin[\mmin(\Phib),\Phib].
\label{phi1b}
\end{equation}
Second, we equate this monomeric volume fraction with the one in our open system
at the saddle point,
\begin{equation}
  \Phimin[\mmax(\Phi),\Phi] = \Phibb(\Phib),
\label{phiopen}
\end{equation}
thus determining (via eqs \ref{mextreme2} and
\ref{phi1extreme}) the total volume fraction in the system,
$\Phi$, and the critical nucleus, $\mmax$, as functions of $\Phib$.
Third, we use these results to calculate $\Phinuc$, $\cnuc$, and $V_1$
as functions of $\Phib$,
\begin{eqnarray}
  \Phinuc(\Phib) &=& \Phi - \Phibb
 \nonumber\\
  \cnuc(\Phib) &=& \Phinuc / (na^3\mmax) \nonumber\\
  V_1(\Phib) &=& \cnuc^{-1} = \frac{na^3\mmax}{\Phi - \Phibb}.
\label{cnucop}
\end{eqnarray}
Finally, the nucleation barrier and time scale are given for the open system by
\begin{eqnarray}
  \Delta F_{\rm nuc}(\Phib) &=& \frac{V_1}{a^3} \left[
  F(\Phi,\Phibb,\mmax) - F_1(\Phibb) \right] \nonumber\\
  \taunuc(\Phib) &\simeq& \tau_0 \rme^{\Delta F_{\rm nuc}(\Phib)}.
\label{taunucop}
\end{eqnarray}

From eqs \ref{F}--\ref{phi1extreme} and
\ref{phi1b}--\ref{taunucop} one can calculate the various
parameters of the nucleation stage for an open system.  Examples are
shown in Figures \ref{fig_cnuc}B and \ref{fig_mmax}B,
revealing striking differences from the case of a closed system. The
explanation is straightforward --- the system is assumed to be in
contact with the reservoir only through its monomeric concentration
(so-called inter-micellar concentration), $\Phibb$, which hardly
changes as $\Phib$ is increased above the {\it cmc}. Hence, during
this initial stage $\Phi_1$ remains low regardless of the value of
$\Phib$. Consequently, the critical nuclei remain relatively rare and
large, almost independent of concentration (Figures
\ref{fig_cnuc}B and \ref{fig_mmax}B).  Moreover, since
$\Phi_1$ does not reach values above $\cC=\cmc$, we get very high
nucleation barriers, resulting in an unphysical nucleation time for
the open system. Thus, homogeneous nucleation in an open system, which
does not have micellar transport from the reservoir, is strongly
hindered. In the following discussion of open systems it is assumed
that, despite this strong kinetic limitation, nuclei were somehow
caused to form.

\section{Micellar Growth}
\label{sec_growth}

The nucleation stage addressed in the preceding section ends when the
critical nuclei have formed. On the free-energy landscape the system
has reached the saddle point of $F$. Subsequently, a stage of faster
growth takes place. The free energy of the system decreases while the
nuclei absorb additional monomers from the surrounding solution and
$m$ increases.

The growth is assumed to be much faster than the nucleation of new
micelles or fusion and fission of existing ones. Hence, the
concentration of micelles, $c_m=(\Phi-\Phi_1)/(na^3m)$ remains fixed
at $c_m=\cnuc$. Consequently, the available volume per aggregate,
$V_1$, remains unchanged as well. We shall assume that the growth is
also faster than the diffusive transport among the micelles (for
closed and open systems) and with the reservoir (open system). The
increase in $m$, therefore, comes solely at the expense of a decrease
in the concentration of the surrounding monomers, while the total
surfactant volume fraction is conserved. Thus, we describe the growth
kinetics as a constrained path, $[\Phi_1(t),m(t)]$, such that
$c_m=\cnuc=\const$ and $\Phi=\const$.

Although diffusive transport into or out of the subsystem (of volume
$V_1$) is assumed negligible during this stage, it is {\it a priori}
unclear whether the growth process itself, within $V_1$, should be
kinetically limited or diffusion-limited. We shall therefore examine
both possibilities below. The constraints and the equations derived in
this section apply to closed and open systems alike, yet the values
substituted for $\Phi$ and $\cnuc$ differ substantially. While for a
closed system $\Phi$ is the experimentally controlled surfactant volume
fraction, for an open system $\Phi$ gets the lower and weakly changing
values determined from $\Phib$ in the nucleation stage according to eq
\ref{phiopen}. The concentration of nuclei is also much lower in the
open-system case (cf.\ \ref{fig_cnuc}). Consequently, the
quantitative predictions for the two cases are quite different.

The aforementioned constraints imply that the average monomer volume
fraction decreases linearly with the aggregation number, $m(t)$,
\begin{equation}
  \Phi_1(t) = \Phi - na^3 \cnuc m(t).
\label{phi1kin}
\end{equation}
We are left with one independent variable, $m(t)$, whose change in
time could be either kinetically controlled or
diffusion-controlled. Yet, before studying the detailed evolution, let
us examine its final state, which is common to both limits.

\begin{figure}
\vspace{0.9cm}
\includegraphics[width=0.48\textwidth]{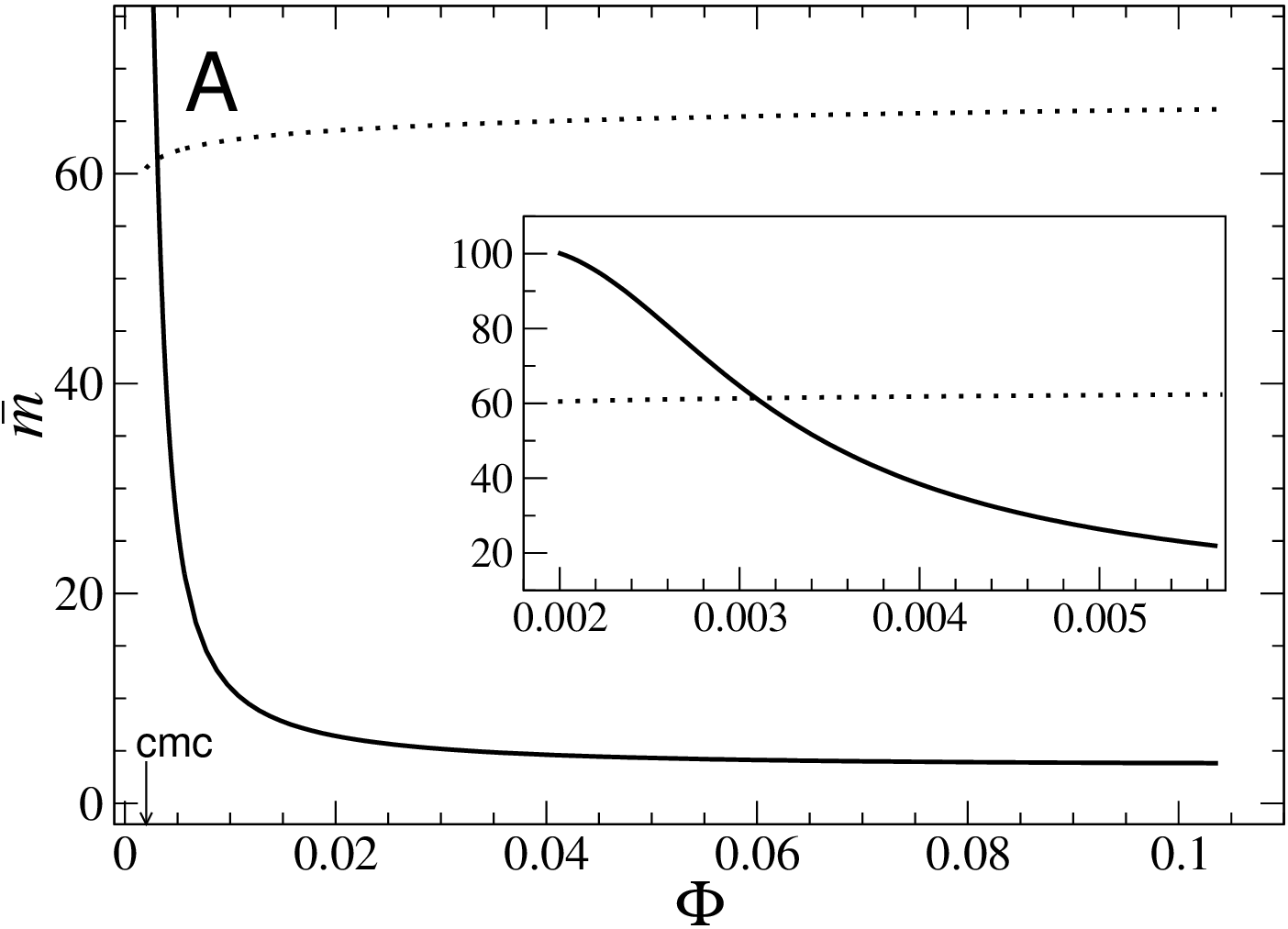}
\hspace{.3cm}
\includegraphics[width=0.48\textwidth]{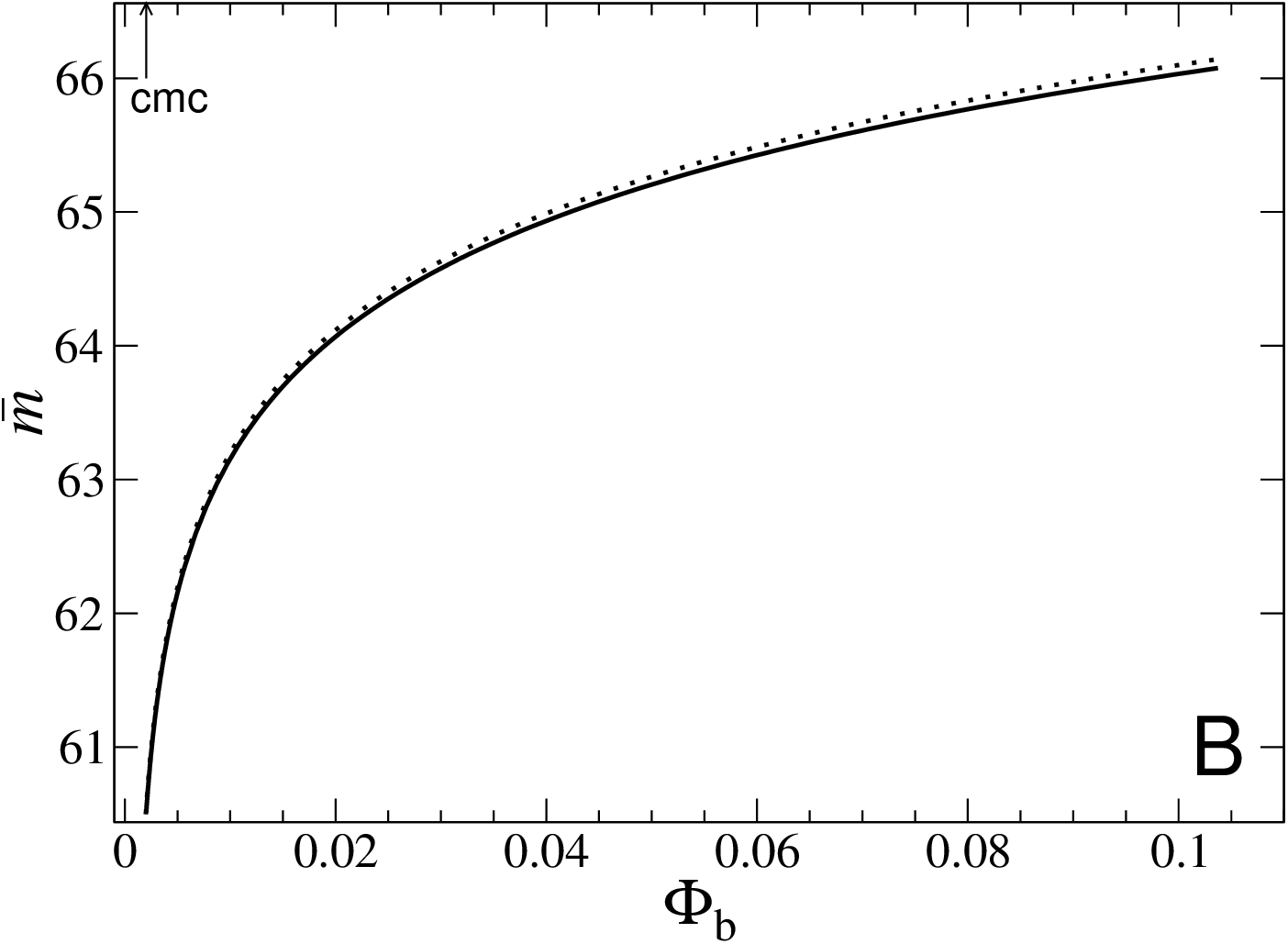}
\caption{Intermediate micelle size at the end of the growth stage,
  $\mkin$, as a function of surfactant volume fraction in the range
  between $\cC=\cmc\simeq 2\times 10^{-3}$ and $\cD$ for a closed (A)
  and open (B) systems. The inset in panel A focuses on volume
  fractions slightly above $\cC$. Dotted lines show the equilibrium
  micelle size, $\mmin$. Parameters are given in
  \ref{tab_surfactant}.}
\label{fig_mkin}
\end{figure}

\begin{figure}
\includegraphics[width=0.47\textwidth]{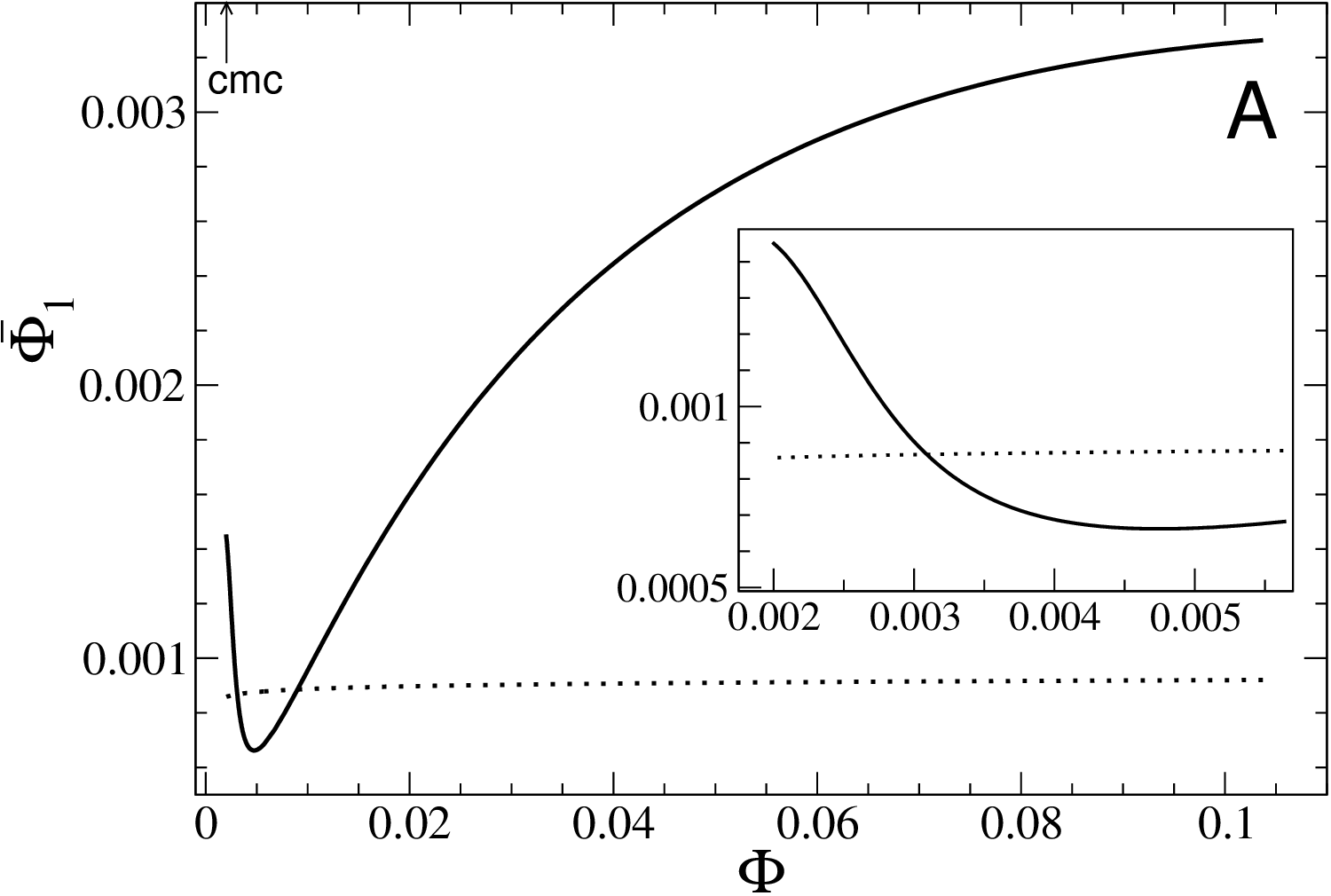}
\hspace{.3cm}
\includegraphics[width=0.48\textwidth]{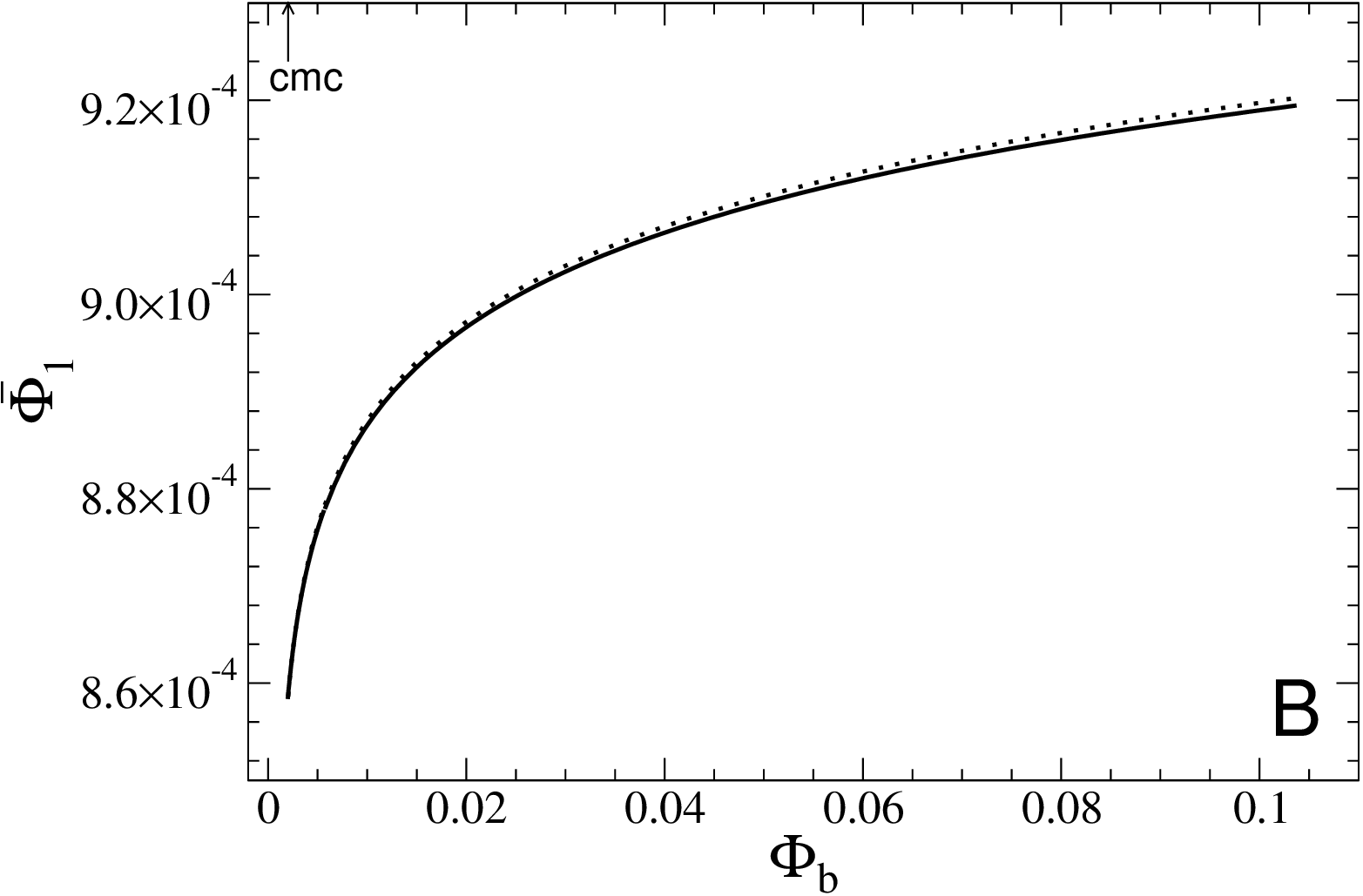}
\caption{Intermediate monomer volume fraction at the end of the growth
  stage as a function of surfactant volume fraction in the range
  between $\cC=\cmc\simeq 2\times 10^{-3}$ and $\cD$ for a closed (A)
  and open (B) systems. The inset in panel A focuses on volume
  fractions slightly above $\cC$.  Dotted lines show for comparison
  the equilibrium monomer volume fraction, $\Phimin$. Parameters are
  given in \ref{tab_surfactant}.}
\label{fig_phi1kin}
\end{figure}

The final state of the growth stage, denoted $(\Phikin,\mkin)$, is
given by the minimum of $F$ along the constrained path, $\left.(\pd F/\pd
m)\right|_{c_m=\cnuc,\Phi={\rm const}}=0$. This yields
\begin{equation}
  m=\mkin:\ \ \ \ln[\Phi_1(m)] + u(m) + mu\,'(m) + 1 - 1/m = 0,
\label{mkin}
\end{equation}
where $\Phi_1(m)$ is given by eq \ref{phi1kin}, and, once $\mkin$
is calculated, $\Phikin=\Phi_1(\mkin)$. The resulting aggregation
numbers and their dependence on the controlled surfactant volume
fraction are presented in \ref{fig_mkin}. Note that the intermediate
aggregation number at the end of the current stage is not equal to the
equilibrium micellar size, since it corresponds to a minimum of $F$
along the constrained path rather than its global minimum.  Unlike the
equilibrium size, $\mmin$, which is bound by thermodynamic stability
to increase with surfactant volume fraction (dotted lines in
\ref{fig_mkin}), the intermediate size $\mkin$ can have a richer
behavior. Examined over a wider range of $\Phi$, $\mkin$ is found to
be nonmonotonous, having a maximum at $\Phi<\cC=\cmc$. Hence, for the
closed system it decreases with $\Phi$ (\ref{fig_mkin}A), whereas for
the open system, which remains dilute throughout this stage, it
increases with $\Phi$ (and, therefore, with $\Phib$; \ref{fig_mkin}B).
In the closed system the growth overshoots the equilibrium size for
$\Phi\gtrsim\cC$ and undershoots it at higher values. Whether $\mkin$
is larger or smaller than $\mmin$ is in accord with the question of
whether $\cnuc$ is, respectively, smaller or larger than the
equilibrium concentration of micelles.  (We shall return to this point
when we deal with the final relaxation.) In the open system $\mkin$ is
very close to, and slightly smaller than, $\mmin$. Similar
observations can be made concerning the intermediate monomer volume
fraction, $\Phikin$, as demonstrated in \ref{fig_phi1kin}.

We now turn to the evolution of the micellar size. We shall first
assume, in the first subsection below, that it is kinetically limited. We
will subsequently check in the second subsection whether such a
description is consistent with the rate of monomer diffusion and
consider the alternative of a diffusive growth.

\subsection{Kinetically Limited Growth}
\label{sec_KLG}

\begin{figure}
\includegraphics[width=0.48\textwidth]{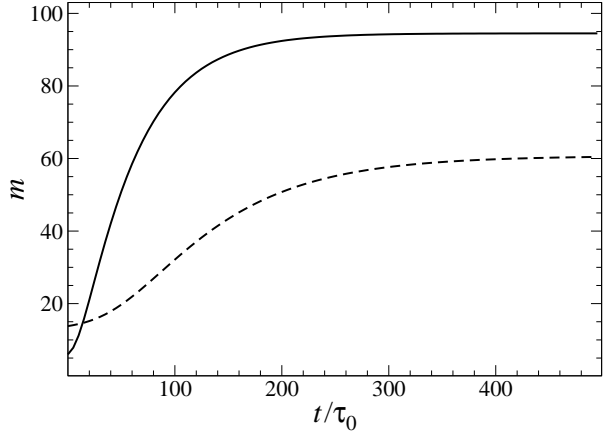}
\caption{Temporal increase in micellar size assuming kinetically
  limited growth in closed (solid line) and open (dashed line)
  systems. The curves are obtained from numerical solution of eq
  \ref{kinetic_eq1} for the parameters given in \ref{tab_surfactant},
  $\alpha=1$, and $\Phi=2.23\times 10^{-3}=1.1\cC$ for the closed
  system, while for the open system the same value is taken for
  $\Phib$.}
\label{mkin_numeric}
\end{figure}

In the case of kinetically limited growth the diffusive transport of
molecules to the aggregate is assumed sufficiently fast so as not to
limit the growth. The volume fraction of monomers, $\Phi_1$, satisfies
eq \ref{phi1kin} while being uniform across the subsystem volume
$V_1$. The increase of $m$ with time is taken as proportional to the
relevant thermodynamic driving force (\ie the slope of $F$ along the
constrained path),
\begin{equation}
  \frac{dm}{dt} = -\frac{\alpha}{\tau_0} \frac{V_1}{a^3}
  \left. \frac{\delta F}{\delta m}
  \right|_{\begin{array}{ll}
  {\scriptstyle c_m=\cnuc}& \\[-12pt]
  {\scriptstyle \Phi={\rm const}}& \end{array}}
  = \frac{\alpha}{\tau_0} \left\{
  \ln[\Phi_1(m)] + u(m) + mu\,'(m) + 1 - 1/m \right\},
\label{kinetic_eq1}
\end{equation}
where $\alpha$ is an unknown dimensionless prefactor of order unity,
and $\Phi_1(m)$ is given by eq \ref{phi1kin}. Equation
\ref{kinetic_eq1}, supplemented by a proper initial condition for
$m(t=0)$, forms a simple initial-value problem for the temporal
increase in micelle size, and is solved numerically. Since the initial
state of this stage is a stationary (saddle) point of $F$, we cannot
begin with the strict initial condition, $m(0)=\mmax$, but have to
perturb it to start the evolution.
An example for a numerical solution of eq \ref{kinetic_eq1}, where we
have taken $m(0)=\mmax+1$ and $\Phi=1.1\cC=1.1\cmc$, is shown in
\ref{mkin_numeric}. The time scale of the growth, denoted $\taukin$,
is found to be about two orders of magnitude larger than the molecular
time $\tau_0$ (\ie of order $10^{-6}$ s in this example).

\begin{figure}
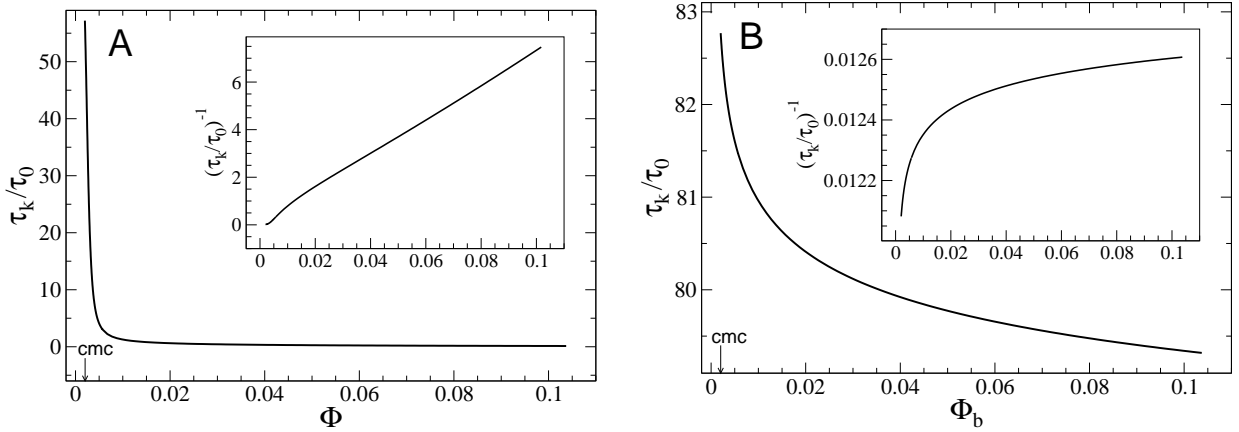

\vspace{0.8cm}
\includegraphics[width=0.48\textwidth]{fig9a.eps}
\hspace{.3cm}
\includegraphics[width=0.48\textwidth]{fig9b.eps}
\caption{Time scale of kinetically limited growth as a function of
  surfactant volume fraction in the range between $\cC=\cmc\simeq
  2\times 10^{-3}$ and $\cD$ for closed (A) and open (B) systems. The
  insets show the increase of $\taukin^{-1}$ (growth rate) with $\Phi$
  (in A) or $\Phib$ (in B).  Parameters are given in
  \ref{tab_surfactant}, and we have set $\alpha=1$ in eq
  \ref{taukin}.}
\label{fig_taukin}
\end{figure}

To get an expression for the kinetic time scale we examine the
asymptotic behavior of eq \ref{kinetic_eq1} as $m$ approaches $\mkin$,
obtaining
\begin{eqnarray}
  &&|m(t)-\mkin| \sim \rme^{-t/\taukin}, \nonumber\\
  &&\taukin^{-1} = \frac{\alpha}{\tau_0} \left[
  \frac{\Phi-\Phi_1}{m\Phi_1} - 2u\,'(m) -mu\,''(m) - 1/m^2
  \right]_{m=\mkin,\Phi_1=\Phikin}.
\label{taukin}
\end{eqnarray}
%
%
The results for $\taukin$ in terms of the molecular time $\tau_0$ are
shown in \ref{fig_taukin}. For the closed system, over one
decade of surfactant volume fraction, $\taukin$ decreases from $\sim
10^2\tau_0$ to $\sim\tau_0$. (Values below $\tau_0$, evidently, should
not be regarded as physical.) The inset shows that the growth rate for
the closed system increases roughly linearly with $\Phi$. For the open
system the time scale is also about two orders of magnitude larger
than $\tau_0$, yet its dependence on $\Phib$ is much weaker for the
reasons described in the Nucleation section.

\subsection{Diffusion-Limited Growth}
\label{sec_DLG}

In the preceding subsection we have assumed that the surrounding
solution can supply the amount of monomers required for micellar
growth within the time scale $\taukin$. Let us check whether this
assumption is consistent with the rate of diffusive transport from the
solution into the aggregate. The thickness of the diffusion layer
around the aggregate, $\ldif$ (assumed to be much larger than the
aggregate radius), satisfies the equation $\Delta m=(4\pi/3)\ldif^3
c_1$, where $\Delta m=\mkin-\mmax$ is the number of monomers to be
transported, and $c_1=\Phi_1/(na^3)$ the monomer concentration. The
diffusion time scale is then $\taudif\sim\ldif^2/D$, $D$ being the
diffusion coefficient of a monomer. Using the definition $\tau_0\sim
(na)^2/D$, we obtain
\begin{equation}
  \taudif/\tau_0 \simeq [3\Delta m/(4\pi n^2)]^{2/3} \Phi_1^{-2/3}
  \sim (\mbox{0.1--1})\Phi_1^{-2/3},
\label{taudif}
\end{equation}
where in the last relation we have assumed $n\sim 10$ and $\Delta
m\sim 50$. For our typical example of $\Phi_1\sim 10^{-3}$ (cf.\
\ref{fig_phi1kin}), we get $\taudif\sim (10$--$10^2)\tau_0$, \ie
comparable to $\taukin$. Thus, the situation concerning the limiting
process for micelle growth is not clearcut, and both processes may be
relevant in general.

To treat the diffusion-limited growth in more detail we employ the
following approximations. First, we neglect the increase in the
aggregate radius, $R$, and take it as constant. Although this
description is evidently inaccurate, it crucially allows us to avoid
the complicated treatment of a moving boundary. Since the growth does
not begin from a single monomer but from a critical nucleus of finite
size $\mmax$, we do not expect the approximation of constant $R$ to
qualitatively affect the results. Second, the diffusion layer is
assumed much smaller than the subsystem, $\ldif\ll V_1^{1/3}$, thus
allowing us to consider the latter as infinite, and the monomer volume
fraction far from the micelle as given by eq \ref{phi1kin}.
Third, we neglect desorption of monomers from the micelle to the
solution during the growth. This is justified in view of the strong
driving force (large slope of $F$) for growth above the
critical-nucleus size.

We assume a radial volume-fraction profile of monomers, $\Phi_1(r>R,t)$,
which follows the diffusion equation,
\begin{equation}
  \frac{\pd\Phi_1}{\pd t} = D \frac{1}{r^2} \frac{\pd}{\pd r}
  \left( r^2 \frac{\pd\Phi_1}{\pd r} \right).
\label{diffusion}
\end{equation}
The growth of a micelle is determined by the diffusive flux of monomers
from the solution,
\begin{equation}
  \frac{dm}{dt} = D \frac{4\pi R^2}{na^3} \left. \frac{\pd\Phi_1}
  {\pd r} \right|_{r=R}.
\label{flux}
\end{equation}
The boundary condition far from the micelle is given according to eq
\ref{phi1kin} by
\begin{equation}
  \Phi_1(r\rightarrow\infty,t) = \Phi - na^3 \cnuc m(t).
\label{bcinfty}
\end{equation}

For the problem to be well posed, eqs
\ref{diffusion}--\ref{bcinfty} should be supplemented by
appropriate initial conditions for $\Phi_1(r,0)$ and $m(0)$, as well
as a local ``adsorption isotherm'' at the aggregate surface, relating
$\Phi_1(R,t)$ and $m(t)$. The latter lies beyond the scope of our
coarse-grained description. At any rate, we are interested primarily
in the qualitative asymptotics of the diffusive transport from the
solution into the aggregate, for which these details are not crucial.
The asymptotic behavior as the final state of the growth stage is
approached is worked out in the Appendix, yielding
\begin{equation}
  \Phi_1(R,t\rightarrow\infty) \simeq \Phikin \left[ 1 - (\taudif/t)^{3/2} \right],
  \ \ \
  \taudif = \frac{a^2 (n\Delta m)^{2/3}}{4\pi D} \Phikin^{-2/3}.
\label{taudifinfty}
\end{equation}
Thus, unlike the exponential relaxation of a kinetically limited
process (eq \ref{taukin}), the diffusive relaxation is
characterized, as usual, by a slow power law. Upon substituting
$\tau_0\sim (na)^2/D$ in eq \ref{taudifinfty} the general form of
$\taudif$, derived earlier from heuristic arguments (eq
\ref{taudif}), is confirmed.

\begin{figure}
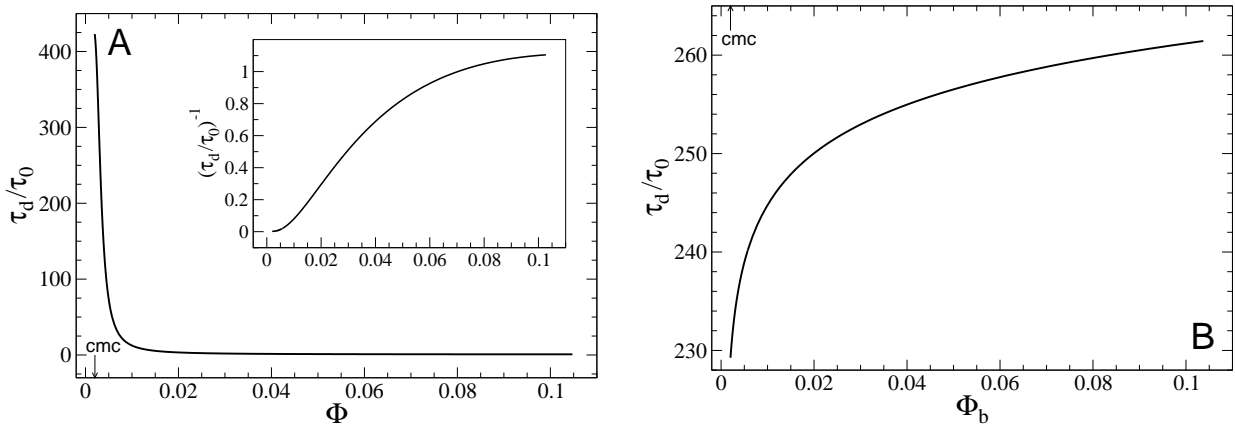

\vspace{0.9cm}
\includegraphics[width=0.48\textwidth]{fig10a.eps}
\hspace{.3cm}
\includegraphics[width=0.48\textwidth]{fig10b.eps}
\caption{Time scale of diffusion-limited growth as a function of
  surfactant volume fraction in the range between $\cC=\cmc\simeq
  2\times 10^{-3}$ and $\cD$ for closed (A) and open (B)
  systems. Parameters are given in \ref{tab_surfactant}.}
\label{fig_taudif}
\end{figure}

\ref{fig_taudif} shows the dependence of $\taudif$ on the controlled
surfactant volume fraction according to eq \ref{taudifinfty},
where we have taken $\tau_0=(na)^2/D$. The cases of closed and open
systems are again found to behave qualitatively differently, $\taudif$
strongly decreasing with $\Phi$ in the former and weakly increasing
with $\Phib$ in the latter. This is a consequence of the different
dependencies of $\mkin$ on concentration, commented on earlier (cf.\
\ref{fig_mkin}). In an open system $\mkin$ increases with $\Phib$
[\ref{fig_mkin}(B)] and, since the more molecules are transported the
longer the diffusive process takes (\ie $\taudif$ increases with
$\Delta m$ in eq \ref{taudifinfty}), we get an increase of $\taudif$
with $\Phib$ [\ref{fig_taudif}(B)].  Comparison of Figures
\ref{fig_taukin} and \ref{fig_taudif} confirms our earlier
assessment, that $\taukin$ and $\taudif$ are comparable in general,
and both growth mechanisms may be relevant.  Only for a closed system
at concentrations slightly above the {\it cmc} do we get for our
representative example $\taudif\gg\taukin$, \ie strictly
diffusion-limited growth. (Note that $\taukin$ and $\taudif$ are
associated with very different time dependencies --- an exponential
law vs.\ a power law --- and are defined only up to a numerical
prefactor. Hence, they should be compared with respect to the order of
magnitude only.)

\subsection{Role of Bulk Diffusion}
\label{sec_bulkD}

In the preceding subsection we have considered the local diffusive
transport that takes place around individual micelles, feeding them
with monomers. In the case of an open system there should also be
slower, long-distance diffusion of monomers from the bulk reservoir.
In principle this should have been the next stage to consider.
However, we find that the monomer volume fraction at the end of the
growth stage, $\Phikin$, is invariably very close to the equilibrium
(bulk) value, $\Phimin$.  (See \ref{fig_phi1kin}B.) This is a
consequence of the small number of initial nuclei (\ref{fig_cnuc}B),
whose growth consumes a small number of monomers. Thus, the driving
force for bulk diffusion is very weak.  Consistently, for the open
system we find also that the micellar size at the end of the growth
stage, $\mkin$, is very close to the equilibrium size, $\mmin$
(\ref{fig_mkin}B).  Therefore, the bulk diffusion that does occur in
an open system has a very minor contribution to the micellization.

\section{Final Relaxation}
\label{sec_relaxation}

At the end of the growth stage monomer transport into the existing
micelles has been exhausted, and the micelles have equilibrated with
the surrounding monomers. Yet, the final state of this stage,
$(\Phikin,\mkin)$, does not correspond to the global free-energy
minimum, since up till now we have constrained the concentration of
micelles to remain at its nucleation value (cf.\ \ref{fig_F}).
A slower process should ensue, therefore, during which the size and/or
concentration of micelles relax to their equilibrium values.

In the open system the situation is a bit unusual. (Recall from the
Nucleation section, however, that actually reaching the current stage
in an open system should already involve overcoming unusually high
barriers.)  The monomer volume fraction has equilibrated with the bulk
reservoir and reached its equilibrium value. The size of the existing
individual micelles has equilibrated as well. What has not
equilibrated yet is the total surfactant volume fraction ---
specifically, the contribution to $\Phi$ from $\Phi_m$, the micellar
volume fraction. Since there is no thermodynamic driving force for
either monomer transport or changes in the size of the existing
micelles, and we do not allow for transport of micelles from the
reservoir, the only open pathway to final relaxation is the very slow
nucleation of additional micelles. The newly formed micelles will take
monomers from the solution, causing transport of additional monomers
from the reservoir, until the total surfactant volume fraction reaches
its equilibrium value, $\Phib$.

The relaxation of the closed system is qualitatively different. Both
the monomer volume fraction and aggregation number have not
equilibrated yet and will change in time while keeping the total
surfactant volume fraction fixed.  Since there is no longer a driving
force for directional exchange of monomers with the solution, we
expect these changes to occur through fusion or fission of micelles.
Such processes occur on the scale of an entire micelle and depend,
therefore, on a different microscopic time, denoted $\tau_m$.  It is
expected to be much larger than the molecular time $\tau_0$ --- either
because of the long diffusion time required for two micelles to meet
before fusing (in which case $\tau_m$ should be of order, say,
$10^{-5}$--$10^{-4}$ s), or due to kinetic barriers for fusion or
fission. In addition, $\tau_m$ should depend on details of
inter-micellar interactions. Such kinetic barriers and interactions
are not accounted for by the current model. Two additional processes,
which in principle could be considered, are irrelevant in this case.
First, nucleation of new micelles or complete disintegration of
existing ones might occur but will require the much longer time scale
of $\taunuc$ discussed earlier.  Second, Ostwald ripening --- a common
relaxation mechanism in phase separation, where larger domains grow at
the expense of smaller ones --- is not expected to take place, since
the finite domains here (the micelles) are not unstable and the
required positive feedback is thus lacking.

Either fission or fusion should be dominant, depending on whether
$\mkin$ has overshot or undershot, respectively, the equilibrium size
$\mmin$. (See \ref{fig_mkin}A.) Correspondingly, the micellar
concentration $c_m$ will either increase or decrease with time. Over
the time scale of these rearrangements of aggregate size and
concentration we can assume that the monomer volume fraction is
relaxed, $\Phi_1(t)=\Phimin[m(t),\Phi]$. We are left again with a
single kinetic variable --- either $m(t)$ or $c_m(t)$.  The two are
related via
\begin{equation}
  c_m(t) = \{\Phi - \Phimin[m(t),\Phi]\}/[na^3m(t)].
\label{c_m}
\end{equation}
The kinetic equation for the micellar size reads
\begin{eqnarray}
  \frac{dm}{dt} &=& -\frac{\beta}{\tau_m} \frac{V_1}{a^3} f(m) \nonumber\\
  f(m) &=& \left.\frac{\delta F}{\delta m}\right|_{\begin{array}{ll} {\scriptstyle
      \Phi_1=\Phimin(m)}&\\[-12pt] {\scriptstyle\Phi={\rm
        const}}& \end{array}} =
  {\Phimin}'\ln\Phimin - \left[ \frac{\Phi-\Phimin}{m^2} + \frac{{\Phimin}'}{m} \right]
  \ln(\Phi-\Phimin)
\label{dmdtrelaxation}\\
  && - (\Phi-\Phimin)u\,'(m) + [u(m)+1-1/m]{\Phimin}', \nonumber
\end{eqnarray}
where $V_1=na^3\mmin/[\Phi-\Phimin(\mmin)]$ is here the volume per
micelle at equilibrium, $\Phimin(m)$ is given by eq \ref{phimin},
a prime denotes $\pd/\pd m$, and $\beta$ is an unknown dimensionless
prefactor of order unity.

\begin{figure}
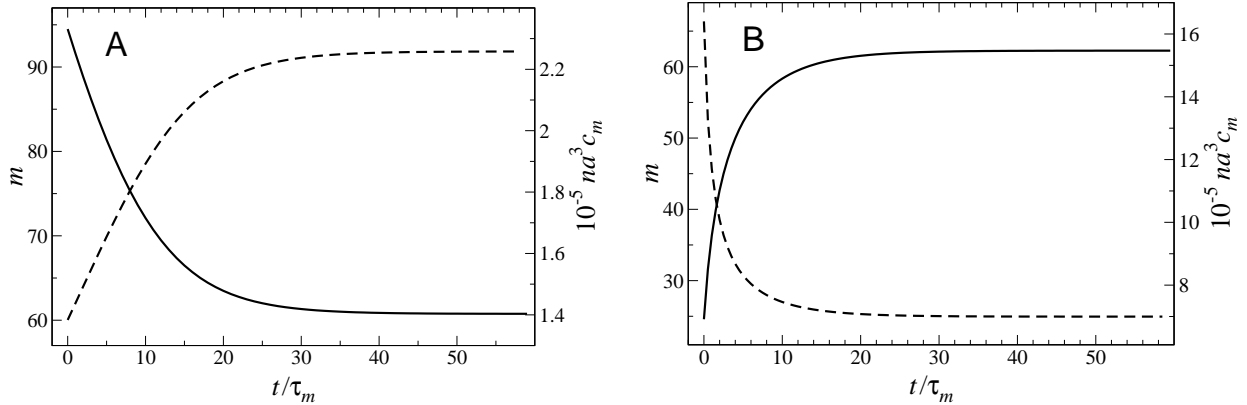

\vspace{0.8cm}
\includegraphics[width=0.48\textwidth]{fig11a.eps}
\hspace{.3cm}
\includegraphics[width=0.48\textwidth]{fig11b.eps}
\caption{Evolution of micellar size (solid, left ordinate) and
  concentration (dashed, right ordinate) during the final relaxation
  stage in a closed system. Parameters are given in
  \ref{tab_surfactant}, we have set $\beta=1$ in eq
  \ref{dmdtrelaxation}, and the volume fraction is
  $\Phi=2.23\times 10^{-3}=1.1\cC$ (A) and $5.23\times
  10^{-3}=2.58\cC$ (B).}
\label{fig_mrel}
\end{figure}

Equations \ref{phimin} and \ref{dmdtrelaxation} are solved
numerically to obtain $m(t)$ and, subsequently (via eq
\ref{c_m}), also $c_m(t)$.  \ref{fig_mrel} shows the solutions
for our exemplary surfactant and two volume fractions, corresponding
to fission- and fusion-dominated relaxation.

\begin{figure}
\vspace{0.9cm}
\includegraphics[width=0.48\textwidth]{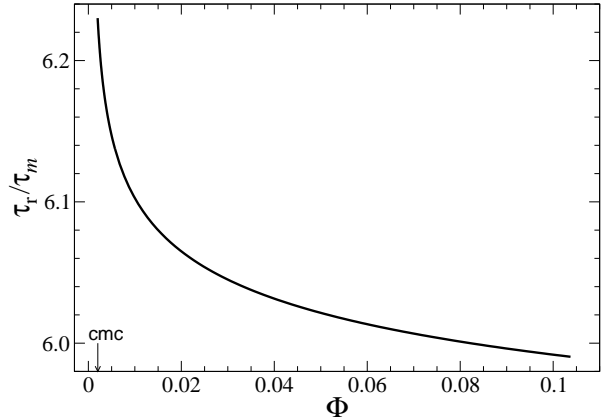}
\caption{Time scale of final relaxation, $\taurel$ (in units of the
  single-micelle time scale), as a function of surfactant volume
  fraction in the range between $\cC=\cmc\simeq 2\times 10^{-3}$ and
  $\cD$ for a closed system.  Parameters are given in
  \ref{tab_surfactant}, and we have set $\beta=1$ in eq
  \ref{taurel}.}
\label{fig_taurel}
\end{figure}

To find the relaxation time we examine the asymptotic behavior of
$m(t\rightarrow\infty)$ according to eq \ref{dmdtrelaxation}, obtaining
\begin{eqnarray}
  &&|m(t) - \mmin| \sim e^{-t/\taurel} \nonumber\\
  &&\taurel = \frac{\tau_m}{\beta} \frac{a^3}{V_1} \frac{1}{f'(\mmin)},
\label{taurel}
\end{eqnarray}
where $f(m)$ has been defined in eq \ref{dmdtrelaxation}. The
dependence of $\taurel$ on surfactant volume fraction is shown in
\ref{fig_taurel}. The relaxation time is found to weakly depend
on $\Phi$, remaining of the same order as (or slightly larger than)
the single-micelle time $\tau_m$ throughout the concentration range.

\section{Discussion}
\label{sec_discuss}

\begin{figure}
\includegraphics[width=0.45\textwidth]{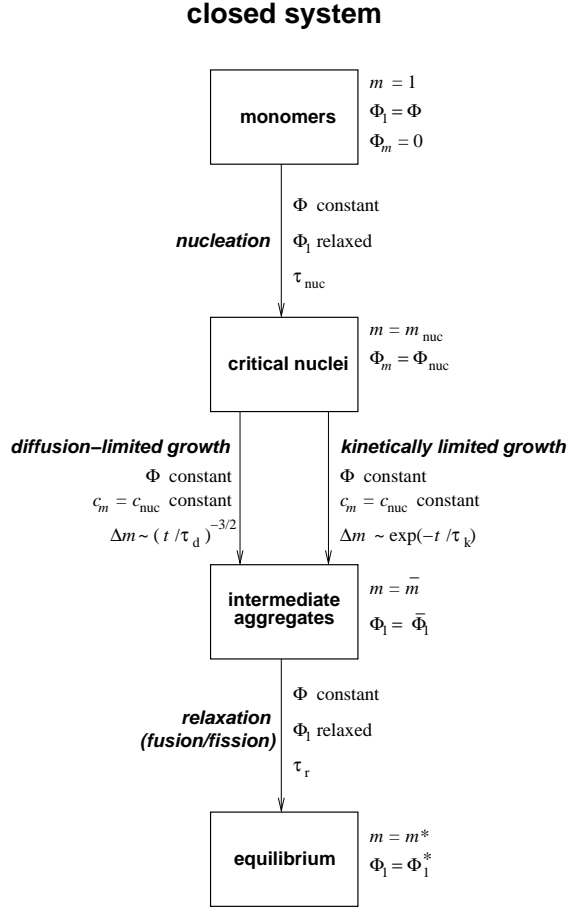}
\caption{Schematic summary of micellization in a closed system. The
  states of the system are represented by rectangles, beside which the
  values of the state variables are indicated. The process is divided
  into three stages, represented by arrows. The constraints on the
  kinetics during each stage are indicated beside the arrow. (i) Slow
  nucleation stage (time scale $\taunuc$), in which critical nuclei of
  size $\mmax$ form in a monomeric solution. (ii) Fast growth stage,
  in which the nuclei grow from $\mmax$ to an intermediate size
  $\mkin$ without changing their concentration. The growth may be
  either diffusion-limited (time scale $\taudif$; $-3/2$ power-law
  relaxation) or kinetically limited (time scale $\taukin$;
  exponential relaxation).  (iii) Final relaxation of the size and
  concentration of aggregates to their equilibrium values through
  fusion or fission (time scale $\taurel$).}
\label{fig_chart1}
\end{figure}

\begin{figure}
\includegraphics[width=0.45\textwidth]{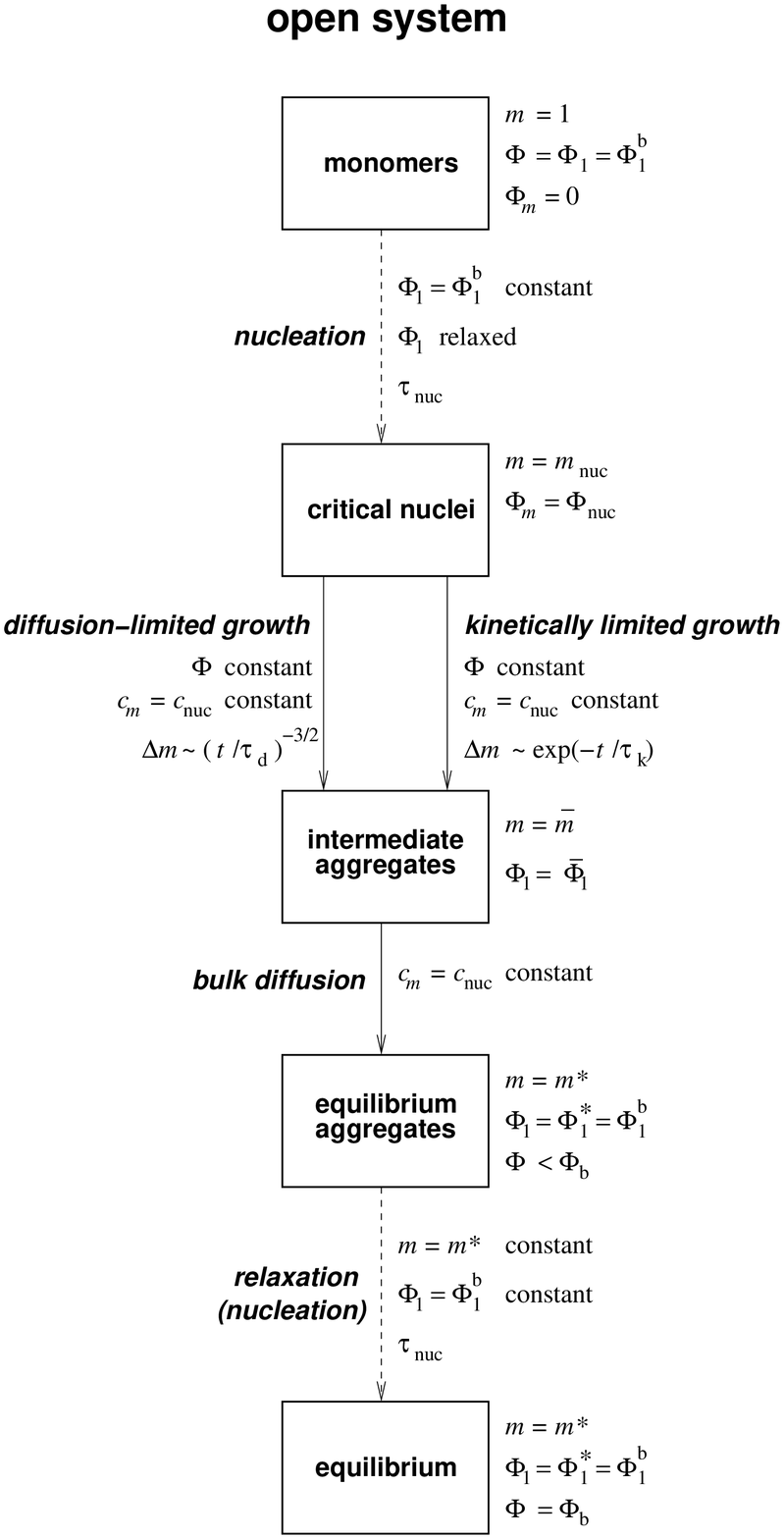}
\caption{Schematic summary of micellization in an open system, having
  monomer exchange with a reservoir. The states of the system are
  represented by rectangles, beside which the values of the state
  variables are indicated. The process is divided into four stages,
  represented by arrows. The constraints on the kinetics during each
  stage are indicated beside the arrow. (i) Slow nucleation stage
  (time scale $\taunuc$), in which critical nuclei of size $\mmax$
  form in a monomeric solution; this stage is found to be strongly
  hindered by kinetic barriers (dashed arrow). (ii) Fast growth stage,
  in which the nuclei grow from $\mmax$ to an intermediate size
  $\mkin$ without changing their concentration. The growth may be
  either diffusion-limited (time scale $\taudif$; $-3/2$ power-law
  relaxation) or kinetically limited (time scale $\taukin$;
  exponential relaxation). (iii) Bulk diffusion from the reservoir
  until the aggregates reach their equilibrium size $\mmin$; this
  stage is found to have a minor contribution to the micellization.
  (iv) Final relaxation of aggregate concentration through nucleation
  of additional micelles (also kinetically hindered; dashed arrow).}
\label{fig_chart2}
\end{figure}

The detailed picture, which arises from our analysis of micellization
kinetics, is schematically summarized in Figures \ref{fig_chart1}
and \ref{fig_chart2}.  We have divided the process of micelle
formation into three major stages --- nucleation, growth, and final
relaxation.  On the one hand, this crude separation into stages should
be conceptually valid, since we find the corresponding time scales to
be quite well separated. In particular, the nucleation time is found
to be macroscopic, several orders of magnitude longer than the time
scales of growth and equilibration. Such three stages have been
resolved in a recent x-ray scattering experiment on block copolymer
micellization \cite{Lund3}.  They also emerged in other micellization
theories \cite{Neu}. On the other hand, the discreteness of these
stages, as illustrated in Figures \ref{fig_chart1} and
\ref{fig_chart2}, should be taken too strictly. In particular, in
the example treated above we find the time scale of growth to be only
$1$--$2$ orders of magnitude shorter than the typical time for final
equilibration. Thus, in certain cases it may well be that these two
stages should not be considered as distinct.

The nucleation stage is much longer than all others and, since it is
an activated process, its duration is exponentially sensitive to
surfactant volume fraction as well as other parameters
(\ref{fig_Fnuc}). The range of nucleation times that we get for our
exemplary surfactant in a closed system (typically larger than $1$ s)
is in line with measured values of $m\tau_2$ --- the time scale for
formation or disintegration of entire micelles \cite{Zana_book}.  The
high nucleation barriers found close to the equilibrium {\it cmc}
($\Phi=\cC$) imply that the measured (apparent) {\it cmc} might in
certain cases be higher than the equilibrium value. This issue, which
was raised before in the context of block copolymer micelles
\cite{Semenov}, clearly merits further study.

The growth stage occurs on much faster time scales (\eg
$10^{-6}$--$10^{-5}$ s for our example). These time scales are similar
to those measured for $\tau_1$ --- the single-monomer exchange time at
equilibrium \cite{Zana_book}. We have found that the growth may in
general be either diffusion-limited or kinetically limited, and that
it should be diffusion-limited at concentrations close to the {\it
  cmc}.  This is in accord with $\tau_1$ being usually
diffusion-limited for short-chain surfactants while becoming
kinetically limited for longer-chain ones, which face higher kinetic
barriers for incorporating into a micelle \cite{Zana_book}. Our theory
predicts a distinctive $-3/2$ power-law relaxation in the case of
diffusion-limited growth (eq \ref{taudifinfty}). This prediction
should be verifiable in scattering experiments like the one described
in ref \citenum{Lund3}, when they are applied to short-chain
surfactants.

The final relaxation stage in a closed system may involve either
reduction in aggregate size (fission), accompanied by an increase in
aggregate concentration, or the other way around (fusion). (See
\ref{fig_mrel}.) Which of these scenarios holds depends on whether the
aggregate size attained in the preceding growth stage has overshot or
undershot the equilibrium aggregation number. The former should hold
at concentrations close to the {\it cmc}, whereas the latter occurs at
higher concentrations. We note that in the experiment of ref
\citenum{Lund3} the aggregates grew in size during their final
relaxation, which is in line with the fact that the amphiphilic
concentration in that experiment was much higher than its {\it cmc}.
We note also that the somewhat surprising possibility of an
intermediate aggregate size overshooting the equilibrium value was
already pointed out in an earlier study \cite{Besseling}. An
interesting consequence of our analysis is that, by tuning to the
right surfactant concentration, one should be able to eliminate the
final-relaxation stage altogether, thus reaching the equilibrium
micellar state already at the end of the fast growth stage. Another
relevant prediction is that the relaxation time of this final stage
should be almost independent of surfactant concentration
(\ref{fig_taurel}). It should be stressed again that these predictions
concerning the final relaxation stage require that the preceding
growth stage be sufficiently fast so that the two processes could be
considered separately. In particular, observing oversized micelles
before they shed their extra molecules may be experimentally
challenging.

Our findings concerning the kinetics of micelle formation have a
number of additional experimental implications. A particularly
clearcut one relates to micellization in an open system --- a solution
in diffusive contact with a reservoir of monomers and micelles.  We
have found that, in cases where only monomer exchange with the
reservoir is allowed while the transport of micelles is blocked,
micellization should be kinetically suppressed. The suppression is
two-fold. First, strong activation is required for the homogeneous
nucleation of the first micelles. This stems from the low surfactant
concentration maintained in the system due to the correspondingly low
monomer concentration (sometimes referred to as the inter-micellar
concentration) in the reservoir.  Second, even after micelles do
nucleate and grow, the final relaxation of their concentration should
be hindered, since it requires the nucleation of additional micelles.

The consequent prediction is that the formation of micelles in such
open monomeric solutions may be suppressed for a macroscopic time.  In
fact, this behavior is regularly manifest in applications involving
micelle-enhanced ultrafiltration \cite{MEUF} and has been observed in
dialysis experiments \cite{Morigaki}, where the time scale of micelle
formation was estimated to be $1$--$10$ hours. In both the
ultrafiltration techniques and the dialysis experiment a micellar
solution is forced through a membrane, whose pores are smaller than
the micelles.  The surfactant solution on the other side of the
membrane remains monomeric for a macroscopic time despite its contact
with a micellar solution above the {\it cmc}. To our best knowledge
the analysis presented above provides the first quantitative account
of this regularly observed behavior.

Apart from the aforementioned strong assumption of time-scale
separation, the main shortcoming of our model is its mean-field
character. We have assumed that the kinetics in the surfactant
solution can be described within a representative subvolume, $V_1$,
containing a single aggregate and being uncorrelated with the other
subvolumes. Upon closer inspection, in fact, we find that $V_1$ for a
closed system typically contains $\sim 10$--$10^2$ surfactant
molecules, which is comparable to the aggregation number. Hence,
correlations among such subvolumes are to be expected as the micelles
nucleate and grow.  Another important mean-field aspect is our
description of the state of the system as a deterministic point on the
free-energy landscape, and its kinetics --- as a sharply defined path
on that landscape. In practice, and particularly close to the {\it
  cmc}, the system should be more accurately described by stochastic
distributions, with polydispersity and occupancies of both the
monomeric and aggregated states \cite{jpcb07}.  Nonetheless, we do not
expect these approximations to qualitative change the main results
presented here.

\section*{acknowledgement}

We are grateful to Raoul Zana and Reidar Lund for helpful discussions.
RH would like to thank Ralf Metzler and the Technical University of Munich for
their hospitality.  Acknowledgment is made to the Donors of the
American Chemical Society Petroleum Research Fund for support of this
research (Grant No.\ 46748-AC6).

\section*{Appendix}
\setcounter{equation}{0}
\renewcommand{\theequation}{A\arabic{equation}}

In this appendix we calculate the asymptotic time dependence of the
micellar size, $m(t)$, in a diffusion-limited growth. The equations to
be handled are eqs \ref{diffusion}--\ref{bcinfty}.

To leading order at long times we can substitute in eq
\ref{bcinfty} $m(t)\simeq\mkin$, turning the boundary condition
far away from the micelle into
$\Phi_1(r\rightarrow\infty,t)=\Phikin$. We now define
$\psi(r,t)=\Phi_1(r,t)-\Phikin$, so that
$\psi(r\rightarrow\infty,t)=0$, and introduce Laplace-transformed
variables, $\psihat(r,s)=\int_0^\infty dt e^{-st}\psi(r,t)$,
$\mhat(s)=\int_0^\infty dt e^{-st} m(t)$. The diffusion equation, eq
\ref{diffusion}, is then rewritten as
\begin{equation}
  s\psihat = D \frac{1}{r^2} \frac{\pd}{\pd r}
  \left( r^2 \frac{\pd\psihat}{\pd r} \right),
\label{L_diffusion}
\end{equation}
where we have assumed $\psi(r,0)=0$, as the accurate initial profile should
not affect the long-time asymptotics. The boundary conditions, eqs
\ref{flux} and \ref{bcinfty}, transform to
\begin{eqnarray}
\label{L_flux}
  &&s\mhat - \mmax = D \frac{4\pi R^2}{na^3}
   \left.\frac{d\psihat}{dr}\right|_{r=R} \\
\label{L_bcinfty}
  &&
   \psihat(r\rightarrow\infty,t) = 0.
\end{eqnarray}

The solution of eqs \ref{L_diffusion}--\ref{L_bcinfty} is
\begin{equation}
  \psihat(r,s) = -\frac{na^3}{4\pi D} \frac{s\mhat-\mmax}{1+R(s/D)^{1/2}}
  \frac{e^{-(s/D)^{1/2}(r-R)}}{r},
\end{equation}
from which we get
\begin{equation}
  \psihat(R,s) = -\frac{na^3}{4\pi DR} \frac{s\mhat-\mmax}{1+R(s/D)^{1/2}}.
\label{psihatR}
\end{equation}
The limit $t\rightarrow\infty$ corresponds to $s\rightarrow 0$, at
which $s\mhat-\mmax\simeq\mkin-\mmax=\Delta m$. Inverting eq
\ref{psihatR} back to real time and taking the limit
$t\rightarrow\infty$, we find
\begin{equation}
  \psi(R,t\rightarrow\infty) \simeq -\frac{na^3\Delta m}{8(\pi Dt)^{3/2}},
\label{psi_R}
\end{equation}
which yields eq \ref{taudifinfty} for $\taudif$.


\end{document}